\documentclass[acmtog, authorversion]{acmart}
\AtBeginDocument{%
  \providecommand\BibTeX{{%
    \normalfont B\kern-0.5em{\scshape i\kern-0.25em b}\kern-0.8em\TeX}}}

\usepackage{amsmath}

\begin{document}

\title{The Future of AI-Assisted Writing}

\author{Carlos Alves Pereira}
\authornote{All authors contributed equally to this research.}
\author{Tanay Komarlu}
\authornotemark[1]
\author{Wael Mobeirek}
\authornotemark[1]
\affiliation{%
  \institution{University of Illinois at Urbana-Champaign}
  \country{USA}
}


\begin{abstract}
The development of Natural Language Generation models has led to the creation of powerful Artificial Intelligence-assisted writing tools. These tools are capable of predicting users' needs and actively providing suggestions as they write. In this work, we conduct a comparative user-study between such tools from an information retrieval lens: pull and push. Specifically, we investigate the user demand of AI-assisted writing, the impact of the two paradigms on quality, ownership of the writing product, and efficiency and enjoyment of the writing process. We also seek to understand the impact of bias of AI-assisted writing. Our findings show that users welcome seamless assistance of AI in their writing. Furthermore, AI helped users to diversify the ideas in their writing while keeping it clear and concise more quickly. Users also enjoyed the collaboration with AI-assisted writing tools and did not feel a lack of ownership. Finally, although participants did not experience bias in our experiments, they still expressed explicit and clear concerns that should be addressed in future AI-assisted writing tools.  
\end{abstract}



\maketitle

\section{Introduction}
Computer-assisted writing tools have been rapidly evolving due to technological advancements. In the modern day, writers have access to spelling and grammar checkers, and online collaborative word processors that have been specifically designed to make writing easier and more efficient for everyone. One variety of writing tools that is gaining traction in the current writing landscape is AI-assisted writing tools. Such tools leverage recent advancements in Artificial Intelligence (AI) and Natural Language Processing (NLP) to enable users to augment their writing through Generative Language Models (GLMs). In fact, AI-assisted writing tools, such as Google Smart Compose (GSC) \cite{GSC}, already exist in a the market in a limited capacity. With these developments, the Human-Computer Interaction aspect of GLMs becomes increasingly important to explore. 

To examine this interaction, we study AI-assisted writing tools from an information retrieval view of access modes: push and pull. We define the push paradigm as the AI choosing when to give suggestions and users accepting them while the pull paradigm is defined as user-requested text generation. The design of these tools changes the way that users undergo the writing process, as they can alleviate the burden of writing from the user and distribute the work to the machine. However, the design of AI-assisted writing tools that successfully fulfill their intended purpose is not a trivial task and raises some important questions. Firstly, we need to understand what users want from these tools? To what extent should these tools be assisting users? Do users want to be provided with a couple of word suggestions, or do users want entire sentences written for them, based on what they've already written? Furthermore, does either of the push or pull paradigms enable users to either produce content more efficiently or improve the quality of their work? Finally, these tools are based on crowdsourced data, so does the generated text show any implicit bias, and is this a problem? \\
In this study, we aim to inform the direction and design of future AI-assisted writing tools. To achieve this goal, we establish our research questions like the following:
\begin{enumerate}
  \item RQ1: To what extent do users want AI to help in the writing process?
  \item RQ2: What is the impact of AI-assisted writing paradigms on the users' sense of ownership and enjoyment?
  \item RQ3: Which AI-assisted writing paradigm enables users to produce higher quality writing more efficiently?
  \item RQ4: How do the biases that can arise through the use of crowdsourced data in AI-assisted writing tools affect users?
\end{enumerate}
To answer these research questions, a comparative experiment was devised that makes use of GSC and GPT-3 to simulate the push and pull paradigms. Users were given two writing tasks. These tasks differed in familiarity to the user, to allow us to analyze how a change in foreknown knowledge and experience in writing a task changes how users use the push and pull paradigms. Users were also only given either the push or pull paradigm, to allow the differences between these results to be measured. They were then surveyed based on their experiences with various quantitative and qualitative questions to describe their experiences. This information was then collated and analyzed to obtain answers to our research questions. We find that users welcome and want help from AI-assisted writing tools. The use of these tools does not negatively impact their sense of ownership and even seemed to enjoy collaboration with the tool. However, the most important point to users is the user interface and user experience of these tools as the interaction should be easy with minimal disruptions. Thus, future writing tools need to carefully design the user experience to be seamlessly integrated into the document editor.

\section{Related Work}
\subsection{Generative Language Models}
NLP has seen drastic advancements in recent years with GLMs. One of the recent GLMs is GPT-3: an autoregressive language model with 175 billion parameters, which is ten times more powerful than any previous non-sparse language model \cite{GPT3Paper}. In all NLP tasks, GPT-3 can be applied without any gradient updates or fine-tuning. The model even excels at tasks that require on-the-fly reasoning or domain adaptation, such as unscrambling words or even utilizing a new word in a sentence. Datasets like Common Crawl which consist of petabytes of data collected over 8 years of web crawling, were included in the GPT-3 training data. Much of GPT-3’s training data was crowdsourced through the use of online text repositories like Wikipedia. As a result, the model can generate samples of news articles that human evaluators have difficulty distinguishing from articles written by humans. These advancements make the possibility of a useful AI-assisted writing tool more reality than fiction. This enables the development of tools that can generate sentences or provide words as suggestions to aid in the writing process.
\subsection{Computer-Assisted Writing}
Research on collaborative computer-assisted writing tools has become more prevalent with GPT-3. For instance, a web application, called SAGA, allows friends to asynchronously collaborate creatively. Through this application, multiple people can contribute to the writing of a story, told partially by GPT-3 \cite{SAGA}. Shakeri et al. explore how to offload the pressure of creative tasks to an AI system. The paper concludes that this allows users to dive deeper into the roleplay experience. However, the paper does not investigate how AI affected the quality of writing. They also did not discuss how the AI impacted the user's sense of enjoyment and satisfaction. It is important to understand if the use of these systems causes users to feel as though they did not create the work. Some users may feel as though the writing pieces are a product of AI. \\
Shelley \cite{Shelley} also develops a GPT-3 based collaborative horror writer that collaboratively writes scary stories with people on Twitter. The application is deployed as a bot on Twitter that generates and posts new stories as tweets. Users can participate in these threads and interact with the stories to produce multiple storylines. To verify whether the generated stories elicit a psychological reaction from readers, Delul et al. utilize measures of effect and anxiety. This work also does not explore the user's sense of ownership and satisfaction. Furthermore, these works do not explore the impact of the possible biases that can arise due to the use of crowd-sourced data. In both papers, the authors explore the application of GPT-3 to develop collaborative writing tools but fail to address whether users want AI help. \\
While these works address the role of the tool to offload creative work, they fail to investigate the use of an AI-assisted writing tool in a professional capacity. Google Smart Compose \cite{GSC} is an AI system for generating interactive, real-time suggestions in Gmail and Google Drive products that assist users in writing tasks by reducing repetitive typing. The system leverages state-of-the-art machine learning techniques for language model training which enables high-quality and accurate suggestion prediction. Chen et al. utilize a large amount of e-mail data to train the neural model at the core of GSC. The aim of the work is to increase the efficiency of a writer by using the local context of the text to provide instant suggestions that complete the sentence as users type.
\subsection{AI-Assisted Writing Paradigms}
We follow the characterizations of AI-Assisted writing paradigms as defined by Clark et al.\cite{MILCreativeWritingClark} In this work, the authors explore the machine-in-the-loop system structure where the loop is initiated with the person providing context and the machine responds with a corresponding suggestion. The person always has control over the final writing output. Clark et al. utilize interaction initiation to represent how a context-suggestion loop is triggered. It can follow a push or pull method of initiation, or a combination of the two. A push condition is defined to be when suggestions are provided to the user while they are writing. On the other hand, a pull condition occurs when the user proactively asks for suggestions. We argue that this view best informs the future of AI-assisted Writing due to the increasing feasibility of using GLMs as an information retrieval tool. This phenomenon is driven by the continuous advancements in NLP as current GLMs reached near human-level performance \cite{GPT3Paper, GSC}.

\begin{table*} 
  \caption{Generative Language Models for AI-assisted Writing Paradigms}
  \label{tab:AI-assisted-writing-models}
  \begin{tabular}{lcc}
    \toprule
    \textbf{Paradigm} & \textbf{Push} & \textbf{Pull}\\
    \midrule
    Model & Google Smart Compose \cite{GSC} & GPT-3 \cite{GPT3Paper} \\
    Training Data & Large corpus of Google emails & 45 TB of text data \\
    Generated Text & Few words & Multiple Sentences\\
    \bottomrule
  \end{tabular}
\end{table*}

\section{Methodology}
\subsection{Writing Task Selection}
The nature of the writing tasks may influence the results of our experiments. To inform our selection, we classify writing into four purposes with related categories:
\begin{itemize}
    \item Writing for Learning: critical, persuasive, and reflective \cite{Dugan2006WritingFC}.
    \item Writing for Academic Communication: scholarly, oral, and technical \cite{Dugan2006WritingFC}.
    \item Writing for Business Communication: team, customer, project management, and career management \cite{Dugan2006WritingFC}.
    \item Writing for Creativity and Entertainment: narrative \cite{SAGA, MILCreativeWritingClark}, and poetic \cite{MILCreativeWritingClark}. 
\end{itemize}

In this work, we focus on Writing for Business Communication as it is the most widely used form of writing. We selected two writing prompts for our experiments. The first prompt asks participants to write a complaint about aging computer hardware. We believe that many participants have experienced frustration from technical issues. As such, they will easily be able to relate their experiences when responding to the prompt. The second prompt asks participants to write an email to request funding for a multi-million dollar building for the participant's university department from a well-known public figure. We suspect that most participants do not have relevant experience, so we provided a few suggested phrases that can be used in the email to kick-start the writing process. We argue that these tasks will allow us to investigate both the pull and push conditions in scenarios where writers have a related experience and in scenarios where a related experience is lacking. Both prompts are written in a scenario-based style to motivate and engage the participants during the writing process. The writing task prompts are included in Appendix \ref{sec:appendix-a}. 

\subsection{Experimental Design}
\subsubsection{User Study Setup}
\begin{figure*}[t]
    \centering
    \includegraphics[width=0.9\textwidth]{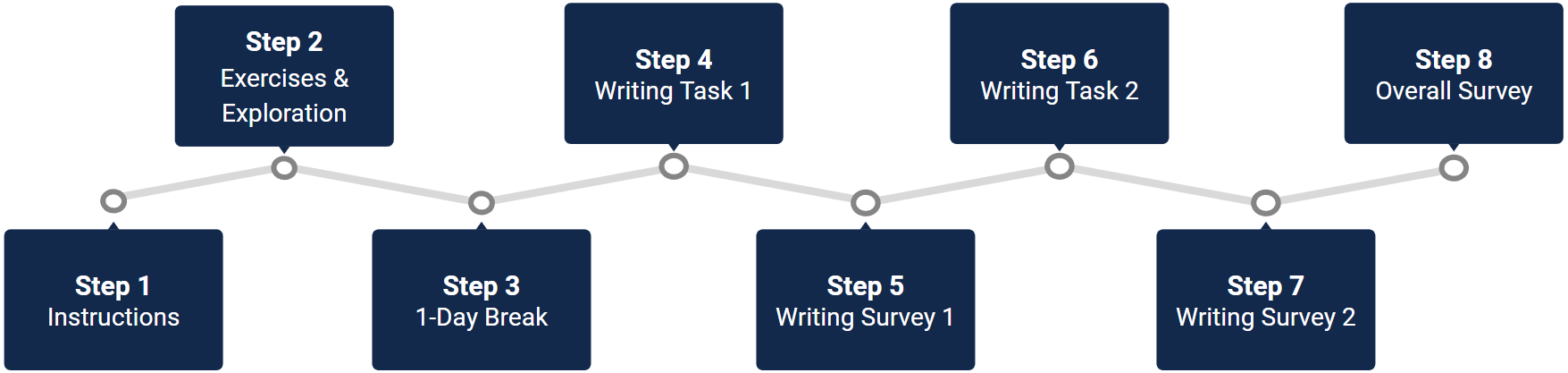}
    \caption{Visualization of the flow of the experiments.}
    \label{fig:exp-design-fig}
\end{figure*}

To answer our research questions, we conduct a user study where participants experience either the push condition or the pull condition. Then, they answer a short survey to reflect on their experience. The experiment flow is divided into 8 steps as visualized in Fig \ref{fig:exp-design-fig}. 

\begin{enumerate}
  \item Instructions Step: Participants are given instructions that explain the overall tasks to perform as well as the tool they will use. The instructions include a screenshot of the tool that allows them to experience the condition they are assigned to.
  \item Exercises and Exploration Step: Once participants indicate that they understood the instructions, they are given a document with short exercises to complete using the tool. After that, participants are freely allowed to explore writing using the tool for their assigned condition. This step addresses the cold-start problem as it allows participants to get more familiar with the tool.
  \item 1-Day Break Step: Participants were given a 1-day break to mitigate the novelty effect of the tools. Participants still had access to the exercise and exploration document during the break.  
  \item Writing Task 1: Participants are given a prompt that sets the scenario and the first writing task. Half of the participants were given the familiar writing task prompt in this step while the other half were given the unfamiliar task prompt.
  \item Writing Task 1 Survey: Participants answer a short survey to reflect on their experience.
  \item Writing Task 2: Participants are given a prompt that sets the scenario and the second writing task. In this step, the participants were given the prompt that they did not complete in step 4. For example, if a participant was given the familiar task in step 4, they would be given the unfamiliar task in this step.
  \item Writing Task Survey 2: Participants answer a short survey to reflect on their experience.
  \item Overall Survey: Participants are asked to answer a short survey that includes some control variables and general reflections about both writing tasks.
\end{enumerate}

\subsubsection{AI-Assisted Writing Tools}
\begin{figure}[t]
    \centering
    \includegraphics[width=9cm]{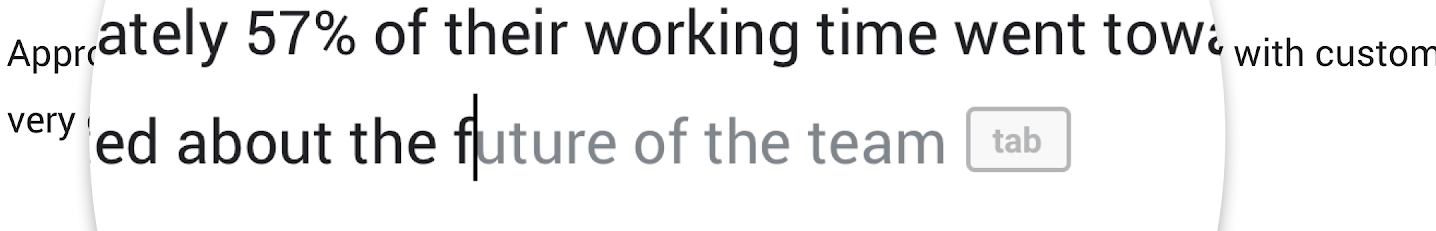}
    \caption{Suggestions by Google Smart Compose in Google Docs}
    \label{fig:gsc-fig}
\end{figure}

To allow participants to experience each condition, we take advantage of Google Docs as its one of the most widely used word processing tools with rich editorial features and support for Add-ons developed by the community. For the push condition, Google Docs allows access to GSC, which suggests text in a lighter color when appropriate. Then, the user can accept the suggestion but pressing tab or can reject the suggestion by continuing to type. At the time of the experiment, GSC within Google Docs is supported for G Suite accounts only, which are typically business-oriented accounts. The interface of Google Docs with a GSC suggestion is shown in Fig \ref{fig:gsc-fig}. 

\begin{figure*}[t!]
    \centering
    \includegraphics[width=0.9\textwidth]{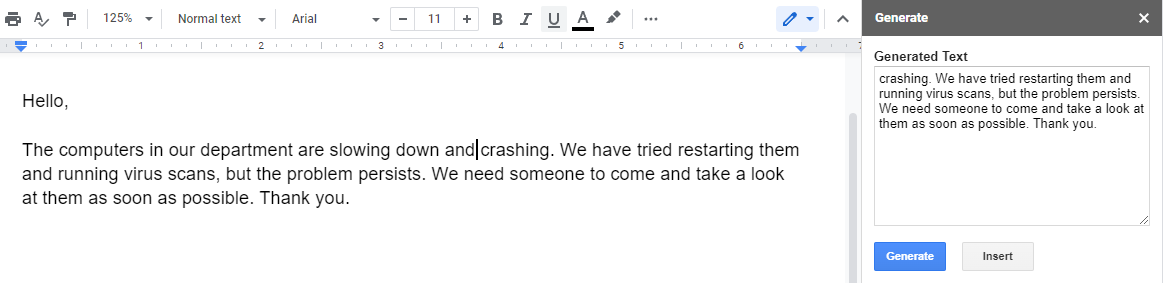}
    \caption{A Google Docs Add-on that generates and inserts text using GPT-3}
    \label{fig:gpt3addon-fig}
\end{figure*}

For the pull condition, we take advantage of the flexibility of Google Docs and developed our own Google Add-on. The Add-on has an interface that allows participants to generate text using GPT-3 by making a call to the OpenAI's open-access application programming interface (API). If the participant is satisfied with the generated text, they can choose to add it to the document using the insert button, The interface of Google Docs with the GPT-3 Add-on is shown in Fig \ref{fig:gpt3addon-fig}. One limitation of the API is that it is difficult to tune to generate a few words that are very likely to be relevant. This can be attributed to the inflexibility of the API as it is not possible to specify the number of sentences generated, only the number of words. This might cause a degradation in the quality of the generated text since half-sentences may not be easily understood by users. As such, we used the default parameters that are recommended by OpenAI.

\subsection{Evaluation}
To evaluate the results of the experiment, we use 5 quantitative metrics.
\subsubsection{User Acceptance}
To measure the user acceptance, we propose the percentage of accepted suggestions defined as follows:
\begin{equation}
    D = \frac{A}{N} \times 100
\end{equation}
where $D$ denotes user acceptance and $A$ denotes the number of accepted suggestions as measured by clicking tab to accept a suggestion for the push condition and the number of clicks on the insert button for the pull condition. $N$ denotes the number of suggestions as measured by the count of suggestions GSC has pushed and the count of clicks on the generate button for the pull condition. For both conditions, we consider the suggestion accepted if the participants insert the suggested text into their document even if it is edited, or removed at a later time point.

\subsubsection{Efficiency} To measure the efficiency, we use a words written per minute metric, defined as follows:
\begin{equation}
\label{e:wpm}
    WPM = \frac{W}{T}
\end{equation}
where $WPM$ denotes efficiency, $W$ denotes the number of words as measured by Google Docs, $T$ denotes the time in minutes that the writing task required to complete.
\begin{table*}[]
\caption{Familiar writing task quantitative results. Higher is better.}
\label{tab:Familiar writing}
\begin{tabular}{lccccc}
\hline
\multicolumn{1}{c}{\textbf{Condition}} & \textbf{Avg. Acceptance} & \textbf{Avg. WPM} & \multicolumn{1}{l}{\textbf{Median Ownership}} & \multicolumn{1}{l}{\textbf{Median Enjoyment}} & \multicolumn{1}{l}{\textbf{Median Quality}} \\ \hline
Pull (N=8)                             & \textbf{54.03}                    & 30.48             & 3.5                                           & 4                                             & \textbf{4.5}                                         \\
Push (N=8)                             & 35.20                    & \textbf{31.55}             & \textbf{5}                                             & \textbf{5}                                             & \textbf{4.5}                                         \\ \hline
\end{tabular}
\end{table*}
\begin{table*}[]
\caption{Unfamiliar writing task quantitative results. Higher is better.}
\label{tab:Unfamiliar writing}
\begin{tabular}{lccccc}
\hline
\multicolumn{1}{c}{\textbf{Condition}} & \textbf{Avg. Acceptance} & \textbf{Avg. WPM} & \multicolumn{1}{l}{\textbf{Median Ownership}} & \multicolumn{1}{l}{\textbf{Median Enjoyment}} & \multicolumn{1}{l}{\textbf{Median Quality}} \\ \hline
Pull (N=8)                             & \textbf{62.50}           & \textbf{54.93}    & 4                                             & 4                                             & 4                                           \\
Push (N=8)                             & 40.90                    & 24.74             & \textbf{4.5}                                  & \textbf{5}                                    & \textbf{4.5}                                \\ \hline
\end{tabular}
\end{table*}
\subsubsection{Quality, Ownership, Enjoyment, and Effect of AI Bias}
To measure the quality of the writing product, the sense of ownership and enjoyment of the writer, and the effect of AI bias on the writer or the writing product, the participants answered five-point Likert scale questions about each metric. The questions were asked for both writing tasks and are identical for both conditions. We included the questions and the scales used in Appendix \ref{sec:appendix-b}.  
\section{Results}
In this section, we perform a quantitative and qualitative analysis of the experimental results. First, we describe participant demographics and control variables. Then, we describe the results relative to each metric we measure.
\subsection{Participant Demographics and Control Variables}
We interviewed 16 participants aged between 18 and 25 years old. All of the participants are college students in the United States majoring in fields related to science, technology, engineering, and mathematics. 3 of the 16 participants spoke English as a second language while the remaining participants were native English speakers. The participants were randomly distributed to experience either the push or the pull conditions. For each condition, half of the participants randomly started with the familiar task, while the other half started with the unfamiliar task. To account for control variables that could affect our experimental results, we asked participants four questions summarized in table \ref{tab:control-variables}. We notice that the median response for all questions is 3, which is the midpoint, except familiarity with the computer issues, which is 4. As such, we can conclude from the results that our experimental design achieved its goals. During the 1-Day break step of the experiment, all participants had access to the tools, but only 2 of the 16 participants used them. 

\begin{table*}[]
\caption{Control Variables Results}
\label{tab:control-variables}
\begin{tabular}{lcc}
\hline
\multicolumn{1}{c}{\textbf{Control Variable}}                                 & \textbf{Median Response} & \textbf{Scale}                                      \\ \hline
Do you actively use AI-assisted writing tools for professional writing tasks? & 3                        & Never - Always \\
How often do you write business emails?                                       & 3                        & Never - Everyday    \\
How familiar are you with experiencing frustrations due to computer issues?   & 4                        & Not at all - All the time                           \\
How familiar are you with fundraising?                                        & 3                        & Not at all - Multiple times                         \\ \hline
\end{tabular}
\end{table*}

\subsection{User Acceptance}
\label{useracceptance}
\begin{figure}[t]
    \centering
    \includegraphics[width=9cm]{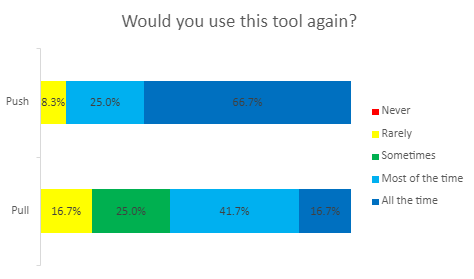}
    \caption{Visualization of continued demand results}
    \label{fig:usetoolagain-fig}
\end{figure}

From the data on the percentage of accepted suggestions in tab. \ref{tab:Familiar writing}, we found that for the pull condition, participants had a higher than 50\% acceptance rate overall, while the push condition had an acceptance rate of at least 35\%. Users were very willing to accept suggestions from AI, and when asked whether or not they would use the tool again in fig. \ref{fig:usetoolagain-fig}, 91.7\% of push participants and 58.3\% of pull participants said they would use the tool most of the time or all of the time. For both the Familiar and Unfamiliar writing tasks, the acceptance of the pull paradigm was significantly higher. Participants in the pull paradigm seemed to use the tool as a method of inspiration, especially when they experienced writer's block. Some Push condition participants indicated that the suggestions provided seemed unnecessary as they were similar to their intended work. Having to actively think about pressing the tab button to accept a suggestion shifted away from their thought process, which they disliked. One advantage of the pull paradigm is the fact that it is on-demand, while the push paradigm gives suggestions unprompted which could disrupt the thought process. Thus, it can lead to the suggestions being inherently ignored as users continue typing. With that said, both push and pull participants have a significant acceptance rate of AI suggestions.

\subsection{Efficiency}
To measure efficiency, we used $WPM$ as described in equation \ref{e:wpm}. The initial expectation is that the pull paradigm would be more efficient as participants would be able to continue writing despite experiencing writer's block using the pull button. Furthermore, GPT-3, which is used to generate text for the pull condition, is likely to generate significantly more words compared to GSC. The unfamiliar task data from our experiments confirm our expectation as the $WPM$ for the pull condition is near twice the push condition $WPM$ as participants experienced writer's block more often. For the familiar task, the difference in $WPM$ seems to be negligible. Participants of the pull condition found that the tool helped them gain insights into fields they did not understand without any external research, which helped in speeding up the writing process. 

\subsection{Quality}
\label{sec:Quality}
\begin{figure}[t]
    \centering
    \includegraphics[width=9cm]{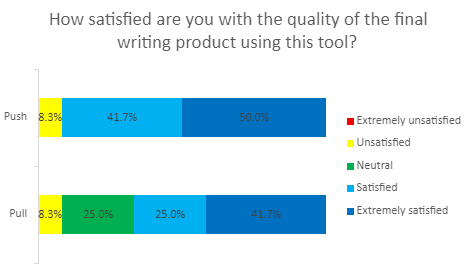}
    \caption{Visualization of quality of writing results}
    \label{fig:quality-fig}
\end{figure}

\begin{figure}[t]
    \centering
    \includegraphics[width=8cm]{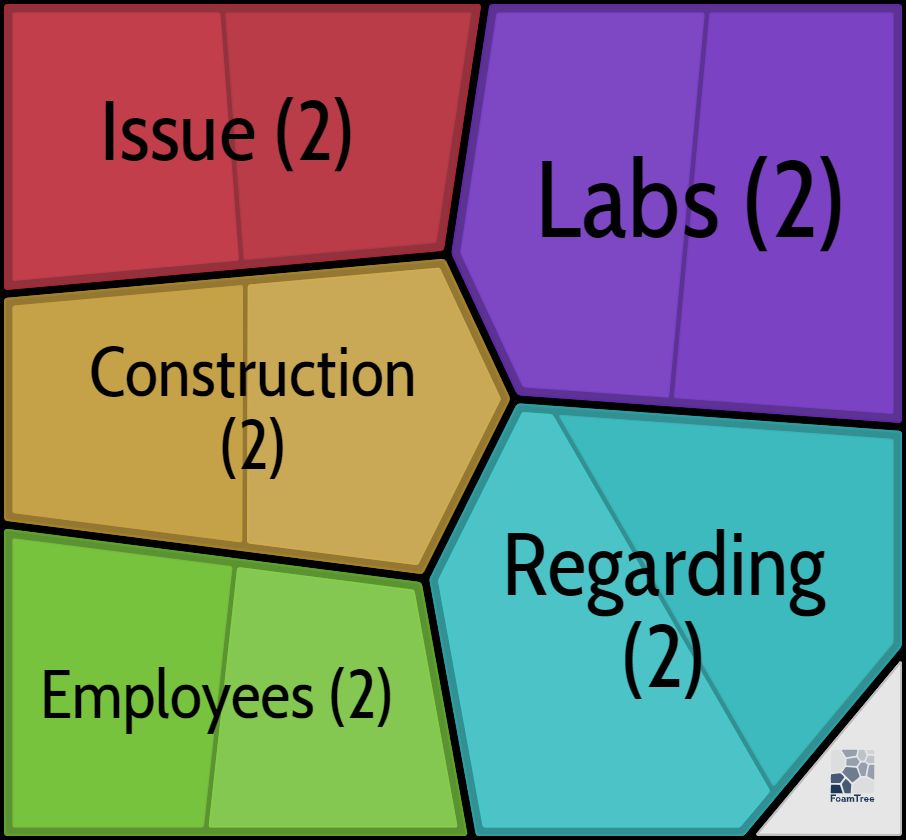}
    \caption{Topic modeling map of the writing of pull condition participants}
    \label{fig:pull-map-fig}
\end{figure}

\begin{figure}[t]
    \centering
    \includegraphics[width=8cm]{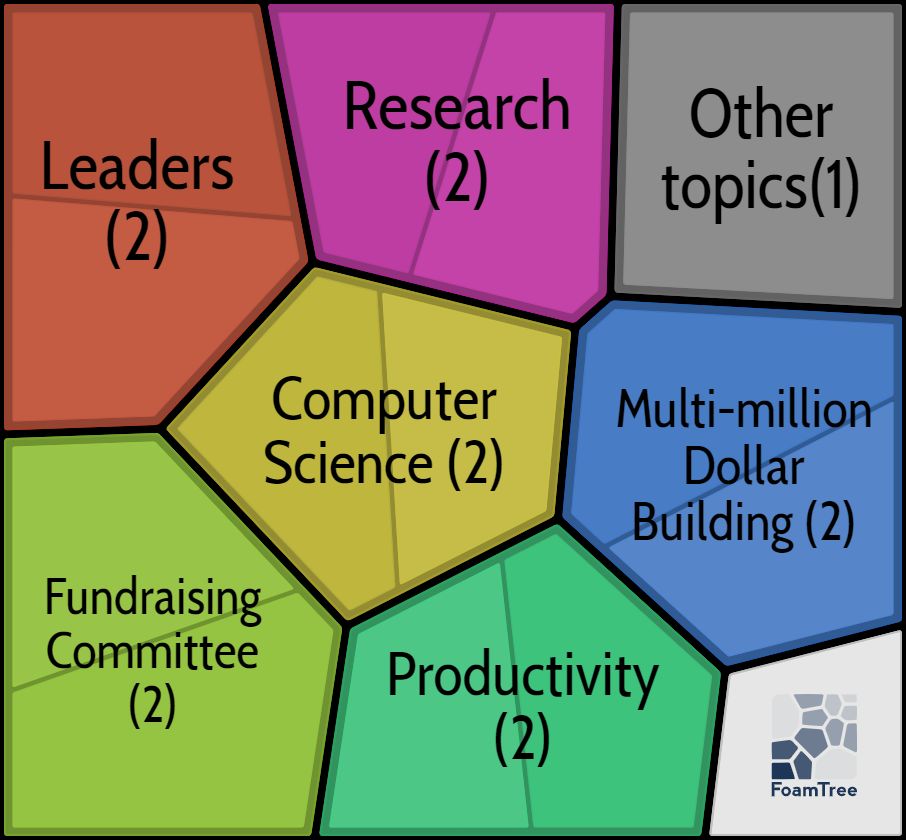}
    \caption{Topic modeling map of the writing of push condition participants}
    \label{fig:push-map-fig}
\end{figure}

In this work, we rely on the participants' judgment to measure the quality of the writing. Our quantitative results are in tab. \ref{tab:Familiar writing} and tab. \ref{tab:Unfamiliar writing} show that participants perceived the quality of the final writing product very similar across both the familiar and unfamiliar writing tasks. This is also confirmed by the distributions in fig. \ref{fig:quality-fig} when comparing the pull and push conditions holistically. From the qualitative data, pull participants mentioned that the tool helped them to gain new ideas and resulted in a more dynamic writing style. This can be seen in fig \ref{fig:pull-map-fig} as the topic modeling results found only five high-level clusters for the pull condition compared to the push's seven clusters in fig. \ref{fig:push-map-fig}. Although more participants are needed for meaningful topic analysis, this shows that pull condition participants included a more diverse set of ideas in their writing. On the other hand, participants in the push condition mention that the tool improved their word choice, grammar, and the overall writing product was more clear and concise. 

\raggedbottom
\subsection{Ownership}
\label{sec:ownership}
Participants found that their sense of ownership was extremely high in the unfamiliar writing task. These users explained that the use of these tools did not impact their sense of ownership as either they ignored the suggestions or they modified the generated text. However, there was a sharp drop in the familiar writing task for the pull paradigm. Some participants found that by pulling multiple sentences, they felt the work was not completely theirs anymore. Furthermore, one user mentioned that they do not attribute the quality of the writing to the tool as they could have produced an email of similar quality without it. With the push paradigm, this may be due to the low acceptance of the suggestions, as participants chose to ignore suggested words from the tool. Participants elaborated that the computer often completed words that they intended to write which allowed them to still feel in control of the writing process. Although ownership may not be as important for business communication, we can still verify our intuition based on the experimental data as it is a relative measure.

\begin{figure}[t]
    \centering
    \includegraphics[width=9cm]{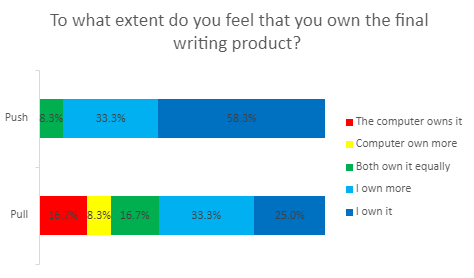}
    \caption{Visualization of sense of ownership results}
    \label{fig:own-fig}
\end{figure}

\subsection{Enjoyment}
\label{sec:enjoyment}
The enjoyment of the participants across both the familiar and unfamiliar tasks was higher in the pull paradigm compared to the push paradigm. This was the opposite of our findings in the pilot study, which did not include a 1-day break with only 4 participants. This suggests that the break was successful in mitigating the novelty effect. A common theme that arose is that users correlate the quality of the writing with the enjoyment of the final writing product. Participants found that the tool provided a good starting point to build off but would not use it as a final piece without modifications. They also indicated that it was entertaining to read the generated text. From our observations, participants were interested in exploring the capabilities of GPT-3 in providing suggestions on multiple topics during step 2, such as philosophy, history, and cooking recipes. Beyond natural language generation, participants explored generating code for programming languages and mathematical equations. However, some participants found that the tool could not provide sufficient context to generate relevant sentences with a single highlight. Furthermore, the enjoyment seemed to wear off after the 1-day break was given. Push condition participants, on the other hand, found that the auto-completion allowed them to make their writing more clear and concise, which they enjoyed.

\begin{figure}[t]
    \centering
    \includegraphics[width=9cm]{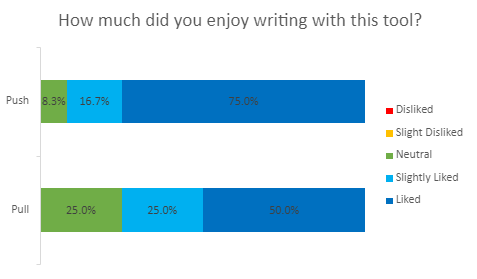}
    \caption{Visualization of enjoyment results}
    \label{fig:enjoy-fig}
\end{figure}

\subsection{Effect of Bias}
\begin{figure}[t]
    \centering
    \includegraphics[width=9cm]{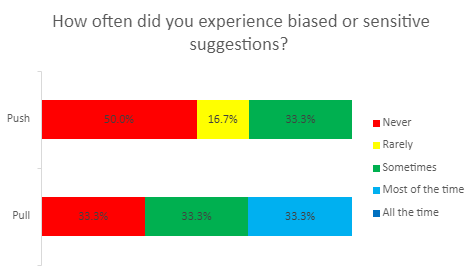}
    \caption{Visualization of frequency of bias results}
    \label{fig:bias-fig}
\end{figure}

The suggestions provided by both of the tools did not have any evidence of biased results based on our observations and qualitative feedback. However, participants voiced concerns regarding the possibility of bias in more complex writing tasks as well as privacy concerns. Although several participants reported some bias as shown in fig. \ref{fig:bias-fig}, we believe that this could be, at least in part, due to the psychological effect of hearing about AI bias and fairness in news and academic outlets as many of the participants are in related fields. A participant even commented that they were impressed by the lack of bias of the generated text. However, one participant found that all of their suggestions were mostly related to academic work and the tool even generated a long tirade from a graduate student.

\section{Discussion}
The results in sec. \ref{useracceptance} suggests that users do want help from AI-assisted writing tools. As expected, users accepted the suggestions more in the unfamiliar task, which is due to experiencing writer's block more often. However, it is interesting to note that the pull condition's acceptance rate is significantly higher across both tasks. This can be due to the main advantage of the pull condition: on-demand suggestions. As such, participants are more likely to be ready to receive and accept suggestions when asking for them. This allowed them access to more ideas that allowed them to continue writing. In comparison, push condition participants felt interrupted by GSC suggestions, which led them to learn to ignore it more often. However, participants preferred the push condition's simplicity and ease of use as it required one button click to insert a suggestion. On the other hand, the pull condition required highlighting text for context in addition to at least 3 button clicks. This shows the importance of designing a seamless workflow for delivering suggestions for a user's writing. Overall, users want AI-assisted writing tools that are well integrated and easy to use.  

To answer RQ2, we consider results from sec. \ref{sec:ownership} and sec. \ref{sec:enjoyment}. Overall, participants felt that they own the writing product and seemed to enjoy AI-assisted writing and across both paradigms and both writing tasks. As expected, participants of the push condition felt more ownership compared to the pull condition since the tool is less intrusive. However, this gap decreases as the task becomes more unfamiliar as we observed participants collaborating with GPT-3 by taking snippets from its suggestions and editing them to achieve their goals. As such, the impact on ownership may not seem as large as we may have expected. Although participants seemed to enjoy GPT-3 initially, the 1-day break was successful in mitigating the novelty effect and participants were less inclined to use it unless they experienced writer's block. Furthermore, push condition participants reported more enjoyment compared to their pull counterparts. We suspect that this is because GSC is a production tool that has been polished over the last few years while the Google Add-on we developed is more clunky and required more clicks. More importantly, however, we believe that the user's readiness for receiving new ideas and suggestions is critical for enjoyment.

For RQ3, we find that participants were equally satisfied with the quality of the writing produced with the AI-assisted writing tools. However, the participants using the pull condition had a much higher average words per minute for the unfamiliar task. This could be attributed to users utilizing the writing tool as a knowledge base to draw ideas. We find that our findings do not conclusively indicate that either the push and pull paradigm enable users to produce higher quality content. We believe that this is due to users modifying the generated text to suit their needs. As a result, many users do not perceive any change in the quality of the writing. To compensate for this effect, we can involve a third-party evaluator to grade the quality of the writings. However, our results do indicate that users prefer that the user experience of the tools be seamlessly integrated into their document editor. 

In addition, we explored the use of crowd-sourced data in AI-assisted writing tools and how the biases that can arise affect users. While many users were concerned about the bias of AI-generated text, most users did not experience any bias. Only 1 in 16 participants experienced biased results. Thus, we conclude that the use of crowd-sourced data should not harm users in most professional use cases. However, the implications of unethical use-cases can be disastrous as many participants expressed concerns about bias and privacy of AI-assisted writing tools, which highlights the importance of this problem.

\section{Conclusion and Future Work}
In conclusion, we find that users do want help from AI-assisted writing tools as it adds to the diversity of ideas and writing styles while maintaining the correctness of grammar and conciseness. Users were not negatively impacted sense of ownership over the final product as because they feel that they have control over the acceptance and rejection of suggestions and even enjoyed the collaboration process. In terms of the quality of the work, both the push and pull paradigms did not impact the quality of the final work. Finally, we also find that the use of crowd-sourced data does not lead to biased results. Many participants do express their concerns regarding the use of this data, which further emphasizes the importance of this question. Future AI-assisted writing tools need to incorporate methods that can detect such content and raise the awareness of users about them. Furthermore, future AI-assisted writing tools need to closely consider their user experience design choices to create a seamless design integrated into the document editor.

Our work can be extended in several ways. First, we would like to improve the user experience design of the Google Add-on for the pull condition to make it more production-ready and comparable to GSC. Secondly, we would like to conduct more experiments across more types of writing purposes, such as writing for learning. Lastly, we would like to conduct an independent expert evaluation to assess the quality of the writing produced with the AI-assisted writing tools.

\bibliographystyle{ACM-Reference-Format}
\bibliography{references}


\begin{thebibliography}{6}


\ifx \showCODEN    \undefined \def \showCODEN     #1{\unskip}     \fi
\ifx \showDOI      \undefined \def \showDOI       #1{#1}\fi
\ifx \showISBNx    \undefined \def \showISBNx     #1{\unskip}     \fi
\ifx \showISBNxiii \undefined \def \showISBNxiii  #1{\unskip}     \fi
\ifx \showISSN     \undefined \def \showISSN      #1{\unskip}     \fi
\ifx \showLCCN     \undefined \def \showLCCN      #1{\unskip}     \fi
\ifx \shownote     \undefined \def \shownote      #1{#1}          \fi
\ifx \showarticletitle \undefined \def \showarticletitle #1{#1}   \fi
\ifx \showURL      \undefined \def \showURL       {\relax}        \fi
\providecommand\bibfield[2]{#2}
\providecommand\bibinfo[2]{#2}
\providecommand\natexlab[1]{#1}
\providecommand\showeprint[2][]{arXiv:#2}

\bibitem[Brown et~al\mbox{.}(2020)]%
        {GPT3Paper}
\bibfield{author}{\bibinfo{person}{Tom~B. Brown}, \bibinfo{person}{Benjamin
  Mann}, \bibinfo{person}{Nick Ryder}, \bibinfo{person}{Melanie Subbiah},
  \bibinfo{person}{Jared Kaplan}, \bibinfo{person}{Prafulla Dhariwal},
  \bibinfo{person}{Arvind Neelakantan}, \bibinfo{person}{Pranav Shyam},
  \bibinfo{person}{Girish Sastry}, \bibinfo{person}{Amanda Askell},
  \bibinfo{person}{Sandhini Agarwal}, \bibinfo{person}{Ariel Herbert-Voss},
  \bibinfo{person}{Gretchen Krueger}, \bibinfo{person}{Tom Henighan},
  \bibinfo{person}{Rewon Child}, \bibinfo{person}{Aditya Ramesh},
  \bibinfo{person}{Daniel~M. Ziegler}, \bibinfo{person}{Jeffrey Wu},
  \bibinfo{person}{Clemens Winter}, \bibinfo{person}{Christopher Hesse},
  \bibinfo{person}{Mark Chen}, \bibinfo{person}{Eric Sigler},
  \bibinfo{person}{Mateusz Litwin}, \bibinfo{person}{Scott Gray},
  \bibinfo{person}{Benjamin Chess}, \bibinfo{person}{Jack Clark},
  \bibinfo{person}{Christopher Berner}, \bibinfo{person}{Sam McCandlish},
  \bibinfo{person}{Alec Radford}, \bibinfo{person}{Ilya Sutskever}, {and}
  \bibinfo{person}{Dario Amodei}.} \bibinfo{year}{2020}\natexlab{}.
\newblock \bibinfo{title}{Language Models are Few-Shot Learners}.
\newblock
\newblock
\urldef\tempurl%
\url{https://doi.org/10.48550/ARXIV.2005.14165}
\showDOI{\tempurl}


\bibitem[Chen et~al\mbox{.}(2019)]%
        {GSC}
\bibfield{author}{\bibinfo{person}{Mia~Xu Chen}, \bibinfo{person}{Benjamin~N.
  Lee}, \bibinfo{person}{Gagan Bansal}, \bibinfo{person}{Yuan Cao},
  \bibinfo{person}{Shuyuan Zhang}, \bibinfo{person}{Justin Lu},
  \bibinfo{person}{Jackie Tsay}, \bibinfo{person}{Yinan Wang},
  \bibinfo{person}{Andrew~M. Dai}, \bibinfo{person}{Zhifeng Chen},
  \bibinfo{person}{Timothy Sohn}, {and} \bibinfo{person}{Yonghui Wu}.}
  \bibinfo{year}{2019}\natexlab{}.
\newblock \showarticletitle{Gmail Smart Compose: Real-Time Assisted Writing}.
\newblock \bibinfo{journal}{\emph{CoRR}}  \bibinfo{volume}{abs/1906.00080}
  (\bibinfo{year}{2019}).
\newblock
\showeprint[arXiv]{1906.00080}
\urldef\tempurl%
\url{http://arxiv.org/abs/1906.00080}
\showURL{%
\tempurl}


\bibitem[Clark et~al\mbox{.}(2018)]%
        {MILCreativeWritingClark}
\bibfield{author}{\bibinfo{person}{Elizabeth Clark},
  \bibinfo{person}{Anne~Spencer Ross}, \bibinfo{person}{Chenhao Tan},
  \bibinfo{person}{Yangfeng Ji}, {and} \bibinfo{person}{Noah~A. Smith}.}
  \bibinfo{year}{2018}\natexlab{}.
\newblock \showarticletitle{Creative Writing with a Machine in the Loop: Case
  Studies on Slogans and Stories}. In \bibinfo{booktitle}{\emph{23rd
  International Conference on Intelligent User Interfaces}} (Tokyo, Japan)
  \emph{(\bibinfo{series}{IUI '18})}. \bibinfo{publisher}{Association for
  Computing Machinery}, \bibinfo{address}{New York, NY, USA},
  \bibinfo{pages}{329–340}.
\newblock
\showISBNx{9781450349451}
\urldef\tempurl%
\url{https://doi.org/10.1145/3172944.3172983}
\showDOI{\tempurl}


\bibitem[Dugan and Polanski(2006)]%
        {Dugan2006WritingFC}
\bibfield{author}{\bibinfo{person}{Robert~F. Dugan} {and}
  \bibinfo{person}{Virginia~G. Polanski}.} \bibinfo{year}{2006}\natexlab{}.
\newblock \showarticletitle{Writing for computer science: a taxonomy of writing
  tasks and general advice}.
\newblock \bibinfo{journal}{\emph{Journal of Computing Sciences in Colleges}}
  \bibinfo{volume}{21} (\bibinfo{year}{2006}), \bibinfo{pages}{191--203}.
\newblock


\bibitem[Shakeri et~al\mbox{.}(2021)]%
        {SAGA}
\bibfield{author}{\bibinfo{person}{Hanieh Shakeri}, \bibinfo{person}{Carman
  Neustaedter}, {and} \bibinfo{person}{Steve DiPaola}.}
  \bibinfo{year}{2021}\natexlab{}.
\newblock \showarticletitle{SAGA: Collaborative Storytelling with GPT-3}. In
  \bibinfo{booktitle}{\emph{Companion Publication of the 2021 Conference on
  Computer Supported Cooperative Work and Social Computing}} (Virtual Event,
  USA) \emph{(\bibinfo{series}{CSCW '21})}. \bibinfo{publisher}{Association for
  Computing Machinery}, \bibinfo{address}{New York, NY, USA},
  \bibinfo{pages}{163–166}.
\newblock
\showISBNx{9781450384797}
\urldef\tempurl%
\url{https://doi.org/10.1145/3462204.3481771}
\showDOI{\tempurl}


\bibitem[Yanardag et~al\mbox{.}(2021)]%
        {Shelley}
\bibfield{author}{\bibinfo{person}{Pinar Yanardag}, \bibinfo{person}{Manuel
  Cebrian}, {and} \bibinfo{person}{Iyad Rahwan}.}
  \bibinfo{year}{2021}\natexlab{}.
\newblock \showarticletitle{Shelley: A Crowd-Sourced Collaborative Horror
  Writer}. In \bibinfo{booktitle}{\emph{Creativity and Cognition}} (Virtual
  Event, Italy) \emph{(\bibinfo{series}{C\&amp;C '21})}.
  \bibinfo{publisher}{Association for Computing Machinery},
  \bibinfo{address}{New York, NY, USA}, Article \bibinfo{articleno}{11},
  \bibinfo{numpages}{8}~pages.
\newblock
\showISBNx{9781450383769}
\urldef\tempurl%
\url{https://doi.org/10.1145/3450741.3465251}
\showDOI{\tempurl}


\end{thebibliography}
\appendix
\section{Writing Prompts}
\label{sec:appendix-a}
\subsection{Familiar Writing Prompt}
Imagine that you are in the office during a typical work day. While you are working, the computer suddenly freezes and shuts down. After a few moments, the computer starts up again, but all of your progress in the last couple of hours is gone. Feeling frustrated, you notice that many of the computers at the office are becoming increasingly slow and old. Write an email to your manager, Mr. Robert Smith, to inform them about the many times the computers overheated, shutdown, and required restarting. (Minimum word count: 100 words)
\subsection{Unfamiliar Writing Prompt}
Imagine you were appointed as a member of the fundraising committee for a brand new, multi-million dollar building at your university for your field of study. The committee's chairman asks you to write an email to Elon Musk asking for a donation for the building. The chairman suggested to use some of the following talking points: offices, labs, lab equipment, meeting rooms, and that the building will have the name of the biggest donor to recognize their contributions. (Minimum word count: 100 words)

\section{Surveys}
\label{sec:appendix-b}
\subsection{Quantitative Questions}
\subsubsection{How satisfied are you with the quality of the final writing product using this tool?}
The range of values were from 1 to 5. The minimum value of 1 was described as 'Extremely unsatisfied' and maximum value of 5 was described as 'Extremely satisfied'.
\subsubsection{How much did you enjoy writing with this tool?}
The range of values were from 1 to 5. The minimum value of 1 was described as 'I disliked writing with this tool' and maximum value of 5 was described as 'I hated writing with this tool'.
\subsubsection{To what extent do you feel that you own the final writing product?}
The range of values were from 1 to 5. The minimum value of 1 was described as 'The computer owns this work' and maximum value of 5 was described as 'I own this work'.
\subsubsection{Would you write using this tool again?}
The range of values were from 1 to 5. The minimum value of 1 was described as 'I would never use this tool again' and maximum value of 5 was described as 'I would use this tool all the time'.
\subsubsection{How familiar are you with experiencing frustration due to computer issues?}
This question was only given for the first writing prompt. The range of values were from 1 to 5. The minimum value of 1 was described as 'Not at all' and maximum value of 5 was described as 'All the time'.
\subsubsection{How familiar are you with fundraising?}
This question was only given for the second writing prompt. The range of values were from 1 to 5. The minimum value of 1 was described as 'Not at all' and maximum value of 5 was described as 'Multiple times'.
\subsubsection{How often do you write business emails?}
The range of values were from 1 to 5. The minimum value of 1 was described as 'I have never written one' and maximum value of 5 was described as 'I write them everyday'.
\subsubsection{Do you actively use AI-assisted writing tools (e.g. Grammarly, Google Smart Compose) for professional writing tasks?}
The range of values were from 1 to 5. The minimum value of 1 was described as 'Never used similar tools' and maximum value of 5 was described as 'Always use similar tools'.
\subsubsection{How often did you experience biased or sensitive suggestions?}
The range of values were from 1 to 5. The minimum value of 1 was described as 'None' and maximum value of 5 was described as 'All the time'.
\subsection{Qualitative Questions}
\subsubsection{Are you satisfied with the final writing product? Why or why not? Please answer with at least 1-2 sentences.}
\subsubsection{How did the use of this tool impact the overall quality of your writing (grammar, coherency, focus, etc)? Please answer with at least 3-4 sentences.}
\subsubsection{How did the use of the tool impact the enjoyment of your writing process? Would you use this tool in the future? Why or why not? Please answer with at least 2-3 sentences.}
\subsubsection{How did the use of this tool impact your sense of ownership? Does it feel like the computer is the sole proprietor of the work? Why or why not? Please answer with at least 2-3 sentences.}
\subsubsection{Did you feel that the suggestions were biased or included sensitive content? How did it impact your experience using the tool? How did it impact your writing? Please answer with at least 2-3 sentences.}
\section{Writing Samples}
\label{sec:appendix-c}
\subsection{Pull Condition Unfamiliar Writing Task Sample}
\textit{Hello Elon,}\\
\textit{I am writing today because the labs at my university could use more funding. The labs at our university are important for students who are pursuing careers in science and engineering. The labs provide students with the opportunity to gain hands-on experience with the equipment and materials they will use in their future careers. However, the labs are in need of more funding in order to maintain their current level of operation.}\\
\textit{I believe that the labs at our university should be given more funding because they are an important resource for students. The labs provide students with the opportunity to gain practical experience with the equipment and materials they will use in their future careers. The labs also help students to develop important problem-solving and critical thinking skills.}\\
\textit{I would appreciate if you could consider giving the labs at our university more funding. Thank you for your time and consideration.}\\
\textit{- Participant 1}
\subsection{Push Condition Familiar Writing Task Sample}
\textit{Hey Robert,}\\
\textit{I am emailing you today for the twenty third time as the computers in the office have gone down again. I know you went on vacation, but I do not have the resources in order to fix this problem. Every time a new computer is booted up, it will start to run windows and then crash. I don’t know if you bought counterfeit windows programs and the computers are rejecting them or if this is a hardware problem. I keep restarting the computers because I have lost all hope and can’t leave work until my hours are complete. Please save me Mr. Robert Smith, you are my only hope. I don’t want to be here anymore. These computers just won’t work.
}\\
\textit{- Participant 2}
\subsection{Push Condition Unfamiliar Writing Task Sample}
\textit{Dear Elon,}\\
\textit{I am writing to you today, to ask for financial support for our new Computer Science building at the University of Illinois Chicago. Not only will your donation help fund our growing program, it will specifically help build labs and meeting areas for our students and faculty. We will also be naming the building after yourself. Since the cost of Computer Science equipment is higher, some of your donations will be allocated to fund high-performance computer equipment. The students in our program will be able to use cutting-edge technology that will be funded by your donation in their research. Thus, making our university’s research program even better. 
}\\
\textit{- Participant 3}
\subsection{Pull Condition Familiar Writing Task Sample}
\textit{Dear Mr.Smith,}\\
\textit{
Hope all is well. I’m writing to inform you regarding issues with the computers in the office. Recently, most of the computers are slow and old. Which also leads to further problems such as overheating, shutting down, and requiring to restart. These issues have led to reduced productivity since it  takes longer to complete tasks.
In order to fix these issues, I suggest that we upgrade the computers in the office. By doing this, it will help with the productivity in the office and hopefully will reduce the amount of problems we are having.
If you have any questions or concerns, please let me know.
Thank you,
}\\
\textit{- Participant 4}
\newpage
\end{document}